\input harvmac
\input epsf
%
\newbox\hdbox%
\newcount\hdrows%
\newcount\multispancount%
\newcount\ncase%
\newcount\ncols
\newcount\nrows%
\newcount\nspan%
\newcount\ntemp%
\newdimen\hdsize%
\newdimen\newhdsize%
\newdimen\parasize%
\newdimen\spreadwidth%
\newdimen\thicksize%
\newdimen\thinsize%
\newdimen\tablewidth%
\newif\ifcentertables%
\newif\ifendsize%
\newif\iffirstrow%
\newif\iftableinfo%
\newtoks\dbt%
\newtoks\hdtks%
\newtoks\savetks%
\newtoks\tableLETtokens%
\newtoks\tabletokens%
\newtoks\widthspec%
%
%
%
%
\tableinfotrue%
\catcode`\@=11
%
%
\def\tstrut{\vrule height3.1ex depth1.2ex width0pt}%
\def\and{\char`\&}
\def\tablerule{\noalign{\hrule height\thinsize depth0pt}}%
\thicksize=1.5pt
\thinsize=0.6pt
\def\thickrule{\noalign{\hrule height\thicksize depth0pt}}%
\def\ctr#1{\hfil\ #1\hfil}%
%
%
%
%
\tablewidth=-\maxdimen%
\spreadwidth=-\maxdimen%
\def\tabskipglue{0pt plus 1fil minus 1fil}%
%
%
\centertablestrue%
%
%
%
%
\parasize=4in%
\gdef\ARGS{########}
\gdef\headerARGS{####}
\def\@mpersand{&}
{\catcode`\|=13
\gdef\letbarzero{\let|0}
\gdef\letbartab{\def|{&&}}%
\gdef\letvbbar{\let\vb|}%
}
{\catcode`\&=4
\def\ampskip{&\omit\hfil&}
\catcode`\&=13
\let&0
\xdef\letampskip{\def&{\ampskip}}%
\gdef\letnovbamp{\let\novb&\let\tab&}
}
\def\begintable{
   \begingroup%
   \catcode`\|=13\letbartab\letvbbar%
   \catcode`\&=13\letampskip\letnovbamp%
   \def\multispan##1{
      \omit \mscount##1%
      \multiply\mscount\tw@\advance\mscount\m@ne%
      \loop\ifnum\mscount>\@ne \sp@n\repeat%
   }
   \def\|{%
      &\omit\widevline&%
   }%
   \ruledtable
}
\long\def\ruledtable#1\endtable{%
%
%
%
   \offinterlineskip
   \tabskip 0pt
   \def\widevline{\vrule width\thicksize}
   \def\endrow{\@mpersand\omit\hfil\crnorm\@mpersand}%
   \def\crthick{\@mpersand\crnorm\thickrule\@mpersand}%
   \def\crthickneg##1{\@mpersand\crnorm\thickrule
          \noalign{{\skip0=##1\vskip-\skip0}}\@mpersand}%
   \def\crnorule{\@mpersand\crnorm\@mpersand}%
   \def\crnoruleneg##1{\@mpersand\crnorm
          \noalign{{\skip0=##1\vskip-\skip0}}\@mpersand}%
   \let\nr=\crnorule
   \def\endtable{\@mpersand\crnorm\thickrule}%
   \let\crnorm=\cr
%
%
   \edef\cr{\@mpersand\crnorm\tablerule\@mpersand}%
   \def\crneg##1{\@mpersand\crnorm\tablerule
          \noalign{{\skip0=##1\vskip-\skip0}}\@mpersand}%
   \let\ctneg=\crthickneg
   \let\nrneg=\crnoruleneg
   \the\tableLETtokens
%
%
   \tabletokens={&#1}
%
%
   \countROWS\tabletokens\into\nrows%
   \countCOLS\tabletokens\into\ncols%
%
%
   \advance\ncols by -1%
   \divide\ncols by 2%
   \advance\nrows by 1%
%
%
   \iftableinfo %
      \immediate\write16{[Nrows=\the\nrows, Ncols=\the\ncols]}%
   \fi%
%
%
   \ifcentertables
      \ifhmode \par\fi
      \line{
      \hss
   \else %
      \hbox{%
   \fi
      \vbox{%
         \makePREAMBLE{\the\ncols}
         \edef\next{\preamble}
         \let\preamble=\next
         \makeTABLE{\preamble}{\tabletokens}
      }
      \ifcentertables \hss}\else }\fi
   \endgroup
   \tablewidth=-\maxdimen
   \spreadwidth=-\maxdimen
}
\def\makeTABLE#1#2{
   {
   \let\ifmath0
   \let\header0
   \let\multispan0
%
%
   \ncase=0%
   \ifdim\tablewidth>-\maxdimen \ncase=1\fi%
   \ifdim\spreadwidth>-\maxdimen \ncase=2\fi%
   \relax
%
   \ifcase\ncase %
      \widthspec={}%
   \or %
      \widthspec=\expandafter{\expandafter t\expandafter o%
                 \the\tablewidth}%
   \else %
      \widthspec=\expandafter{\expandafter s\expandafter p\expandafter r%
                 \expandafter e\expandafter a\expandafter d%
                 \the\spreadwidth}%
   \fi %
   \xdef\next{
      \halign\the\widthspec{%
      #1
      \noalign{\hrule height\thicksize depth0pt}
      \the#2\endtable
%
      }
   }
   }
   \next
}
\def\makePREAMBLE#1{
   \ncols=#1
   \begingroup
   \let\ARGS=0
   \edef\xtp{\widevline\ARGS\tabskip\tabskipglue%
   &\ctr{\ARGS}\tstrut}
   \advance\ncols by -1
   \loop
      \ifnum\ncols>0 %
      \advance\ncols by -1%
      \edef\xtp{\xtp&\vrule width\thinsize\ARGS&\ctr{\ARGS}}%
   \repeat
   \xdef\preamble{\xtp&\widevline\ARGS\tabskip0pt%
   \crnorm}
   \endgroup
}
\def\countROWS#1\into#2{
   \let\countREGISTER=#2%
   \countREGISTER=0%
   \expandafter\ROWcount\the#1\endcount%
}%
\def\ROWcount{%
   \afterassignment\subROWcount\let\next= %
}%
\def\subROWcount{%
   \ifx\next\endcount %
      \let\next=\relax%
   \else%
      \ncase=0%
      \ifx\next\cr %
         \global\advance\countREGISTER by 1%
         \ncase=0%
      \fi%
      \ifx\next\endrow %
         \global\advance\countREGISTER by 1%
         \ncase=0%
      \fi%
      \ifx\next\crthick %
         \global\advance\countREGISTER by 1%
         \ncase=0%
      \fi%
      \ifx\next\crnorule %
         \global\advance\countREGISTER by 1%
         \ncase=0%
      \fi%
      \ifx\next\crthickneg %
         \global\advance\countREGISTER by 1%
         \ncase=0%
      \fi%
      \ifx\next\crnoruleneg %
         \global\advance\countREGISTER by 1%
         \ncase=0%
      \fi%
      \ifx\next\crneg %
         \global\advance\countREGISTER by 1%
         \ncase=0%
      \fi%
      \ifx\next\header %
         \ncase=1%
      \fi%
      \relax%
      \ifcase\ncase %
         \let\next\ROWcount%
      \or %
         \let\next\argROWskip%
      \else %
      \fi%
   \fi%
   \next%
}
\def\counthdROWS#1\into#2{%
\dvr{10}%
   \let\countREGISTER=#2%
   \countREGISTER=0%
\dvr{11}%
\dvr{13}%
   \expandafter\hdROWcount\the#1\endcount%
\dvr{12}%
}%
\def\hdROWcount{%
   \afterassignment\subhdROWcount\let\next= %
}%
\def\subhdROWcount{%
   \ifx\next\endcount %
      \let\next=\relax%
   \else%
      \ncase=0%
      \ifx\next\cr %
         \global\advance\countREGISTER by 1%
         \ncase=0%
      \fi%
      \ifx\next\endrow %
         \global\advance\countREGISTER by 1%
         \ncase=0%
      \fi%
      \ifx\next\crthick %
         \global\advance\countREGISTER by 1%
         \ncase=0%
      \fi%
      \ifx\next\crnorule %
         \global\advance\countREGISTER by 1%
         \ncase=0%
      \fi%
      \ifx\next\header %
         \ncase=1%
      \fi%
\relax%
      \ifcase\ncase %
         \let\next\hdROWcount%
      \or%
         \let\next\arghdROWskip%
      \else %
      \fi%
   \fi%
   \next%
}%
{\catcode`\|=13\letbartab
\gdef\countCOLS#1\into#2{%
   \let\countREGISTER=#2%
   \global\countREGISTER=0%
   \global\multispancount=0%
   \global\firstrowtrue
   \expandafter\COLcount\the#1\endcount%
   \global\advance\countREGISTER by 3%
   \global\advance\countREGISTER by -\multispancount
}%
\gdef\COLcount{%
   \afterassignment\subCOLcount\let\next= %
}%
{\catcode`\&=13%
\gdef\subCOLcount{%
   \ifx\next\endcount %
      \let\next=\relax%
   \else%
      \ncase=0%
      \iffirstrow
         \ifx\next& %
            \global\advance\countREGISTER by 2%
            \ncase=0%
         \fi%
         \ifx\next\span %
            \global\advance\countREGISTER by 1%
            \ncase=0%
         \fi%
         \ifx\next| %
            \global\advance\countREGISTER by 2%
            \ncase=0%
         \fi
         \ifx\next\|
            \global\advance\countREGISTER by 2%
            \ncase=0%
         \fi
         \ifx\next\multispan
            \ncase=1%
            \global\advance\multispancount by 1%
         \fi
         \ifx\next\header
            \ncase=2%
         \fi
         \ifx\next\cr       \global\firstrowfalse \fi
         \ifx\next\endrow   \global\firstrowfalse \fi
         \ifx\next\crthick  \global\firstrowfalse \fi
         \ifx\next\crnorule \global\firstrowfalse \fi
         \ifx\next\crnoruleneg \global\firstrowfalse \fi
         \ifx\next\crthickneg  \global\firstrowfalse \fi
         \ifx\next\crneg       \global\firstrowfalse \fi
      \fi
\relax
      \ifcase\ncase %
         \let\next\COLcount%
      \or %
         \let\next\spancount%
      \or %
         \let\next\argCOLskip%
      \else %
      \fi %
   \fi%
   \next%
}%
\gdef\argROWskip#1{%
   \let\next\ROWcount \next%
}
\gdef\arghdROWskip#1{%
   \let\next\ROWcount \next%
}
\gdef\argCOLskip#1{%
   \let\next\COLcount \next%
}
}
}
\def\spancount#1{
   \nspan=#1\multiply\nspan by 2\advance\nspan by -1%
   \global\advance \countREGISTER by \nspan
   \let\next\COLcount \next}%
\def\dvr#1{\relax}%
\def\header#1{%
\dvr{1}{\let\cr=\@mpersand%
\hdtks={#1}%
\counthdROWS\hdtks\into\hdrows%
\advance\hdrows by 1%
\ifnum\hdrows=0 \hdrows=1 \fi%
\dvr{5}\makehdPREAMBLE{\the\hdrows}%
\dvr{6}\getHDdimen{#1}%
{\parindent=0pt\hsize=\hdsize{\let\ifmath0%
\xdef\next{\valign{\headerpreamble #1\crnorm}}}\dvr{7}\next\dvr{8}%
}%
}\dvr{2}}
\def\makehdPREAMBLE#1{
\dvr{3}%
\hdrows=#1
{
\let\headerARGS=0%
\let\cr=\crnorm%
\edef\xtp{\vfil\hfil\hbox{\headerARGS}\hfil\vfil}%
\advance\hdrows by -1
\loop
\ifnum\hdrows>0%
\advance\hdrows by -1%
\edef\xtp{\xtp&\vfil\hfil\hbox{\headerARGS}\hfil\vfil}%
\repeat%
\xdef\headerpreamble{\xtp\crcr}%
}
\dvr{4}}
\def\getHDdimen#1{%
\hdsize=0pt%
\getsize#1\cr\end\cr%
}
\def\getsize#1\cr{%
\endsizefalse\savetks={#1}%
\expandafter\lookend\the\savetks\cr%
\relax \ifendsize \let\next\relax \else%
\setbox\hdbox=\hbox{#1}\newhdsize=1.0\wd\hdbox%
\ifdim\newhdsize>\hdsize \hdsize=\newhdsize \fi%
\let\next\getsize \fi%
\next%
}%
\def\lookend{\afterassignment\sublookend\let\looknext= }%
\def\sublookend{\relax%
\ifx\looknext\cr %
\let\looknext\relax \else %
   \relax
   \ifx\looknext\end \global\endsizetrue \fi%
   \let\looknext=\lookend%
    \fi \looknext%
}%
%
%
\def\tablelet#1{%
   \tableLETtokens=\expandafter{\the\tableLETtokens #1}%
}%
\catcode`\@=12

\def\frak#1#2{{\textstyle{{#1}\over{#2}}}}
\def\frakk#1#2{{{#1}\over{#2}}}
\def\fivebar{{\overline{5}}}
\def\tenbar{{\overline{10}}}

\def\ttilde{\tilde t}
\def\utilde{\tilde u}

\def\sic{supersymmetric}

\def\npb{{Nucl.\ Phys.\ }{\bf B}}
\def\plb{{Phys.\ Lett.\ }{ \bf B}}

\def\prd{{Phys.\ Rev.\ }{\bf D}}

\def\lf{16\pi^2}

\def\TeV{{\rm TeV}}
\def\GeV{{\rm GeV}}

\thicksize=0.7pt
\thinsize=0.5pt
\def\ctr#1{\hfil $\,\,\,#1\,\,\,$ \hfil}
\def\tstrut{\vrule height 2.7ex depth 1.0ex width 0pt}

\def \inparg{\leftskip = 40pt\rightskip = 40pt}
\def \outparg{\leftskip = 0 pt\rightskip = 0pt}

{\nopagenumbers
\line{\hfil LTH 587 }
\line{\hfil hep-ph/0308231}
\line{\hfil Revised Version}
\vskip .5in
\centerline{\titlefont Three loop soft running, benchmark points} 
\medskip
\centerline{\titlefont and semi-perturbative unification}
\vskip 1in
\centerline{\bf I.~Jack, D.R.T.~Jones\foot{address from Sept 1st 2003-
31 Aug 2004: TH Division, CERN, 1211 Geneva 23, Switzerland} and A.F.~Kord}
\bigskip
\centerline{\it Department of Mathematical Sciences, 
University of Liverpool, Liverpool L69 3BX, U.K.}
\vskip .3in

We consider three-loop $\beta$-function corrections to 
the sparticle spectrum in the MSSM, with particular emphasis on  Snowmass 
Benchmark points. The three loop running has little effect on 
the weakly interacting particle spectrum,  but for the squark masses 
the  effect can be comparable to, or greater than, that of 
two loop running. We extend the analysis to the 
semi-perturbative unification scenario, where the impact of the 
three loop corrections becomes even more dramatic.

\Date{August 2003}}  

\newsec{Introduction}

Softly broken supersymmetry remains a well motivated and popular 
playground for Beyond the Standard Model practitioners. 
Calculations of sparticle spectra resulting from given assumptions 
about the underlying theory have become increasingly refined, with several 
public programs available that incorporate two-loop Renormalisation Group 
Equations (RGEs) and one-loop radiative corrections. For a recent comparison 
of the output of some of these programs   
see the paper by Allanach, Kraml and Porod (AKP) \ref\AllanachJW{
B.C.~Allanach, S.~Kraml and W.~Porod,
JHEP  016 (2003) 0303
}. Two loop corrections are also available to, for example, the 
effective potential
\ref\MartinIU{
S.P.~Martin, \prd66 (2002) 096001\semi
A.~Dedes, G.~Degrassi and P.~Slavich,
hep-ph/0305127\semi
A.~Dedes and P.~Slavich,
\npb657 (2003) 333 
}.
One area where there has been considerable progress in the last few 
years is in the calculation of the RGE $\beta$-functions. For an
arbitrary \sic\ theory the chiral supermultiplet anomalous dimension 
$\gamma$ is 
known to three loops\ref\JackQQ{
I.~Jack, D.R.T.~Jones and C.G.~North,
\npb473 (1996) 308
}
and the gauge $\beta$-function(s) $\beta_g$ to four loops 
\ref\JackCN{
I.~Jack, D.R.T.~Jones and C.G.~North,
\npb486 (1997) 479 
}. For the MSSM, 
both $\beta_{g_i}$ and the various $\gamma$s were given through 
three loops in Ref.~\ref\fjja{P.M.~Ferreira, I.~Jack and D.R.T.~Jones,
\plb387 (1996) 80 
}.

The $\beta$-functions in a general theory and the MSSM 
for the ``standard '' soft breaking terms 
were given to two loops in 
Ref.~\ref\mv{
I.~Jack and
D.R.T.~Jones, \plb333 (1994) 372 \semi
S.P.~Martin
and M.T.~Vaughn, \prd50 (1994) 2282  \semi
Y.~Yamada, \prd50 (1994) 3537 }
and for the ``non-standard'' terms   
also to two loops in Ref.~\ref\jjns{I.~Jack and D.R.T.~Jones, \plb 457 (1999)
101}. Recently, however, it has been 
realised\ref\jjpa{I.~Jack and   D.R.T.~Jones 
\plb415 (1997) 383 }%
\nref\jjpb{I.~Jack, D.R.T.~Jones and A.~Pickering,  
\plb432 (1997) 114}
--\ref\akk{L.V.~Avdeev, D.I.~Kazakov and I.N.~Kondrashuk, 
\npb510 (1998) 289}\ that in the case 
of the ``standard'' terms it is possible to express the associated 
$\beta$-functions exactly in terms of simple differential  operators 
acting on $\beta_{g_i}$ and $\gamma$. It is therefore a straightforward 
matter to derive the three-loop ``standard'' soft $\beta$-functions 
for the MSSM and variations thereof. In this paper we present  
three-loop running results for the MSSM with the 
addition of $n_5$ and $n_{10}$ sets of $SU_5$  $5(\fivebar)$ and $10(\tenbar)$
representations respectively. A motive for grouping the additional 
matter in this way is that complete $SU_5$ representations do not 
(at one loop) change the prediction of $\sin^2\theta_W$ (or alternatively 
of $g_3^2 (M_Z)$)  that follows 
from imposing $g_{1,2,3}$ gauge unification. Also unchanged 
at one loop is the gauge unification scale, $M_X$; 
but at higher loops this scale increases and can 
approach the string scale. (For a recent account of unification 
at the string scale via the addition of {\it incomplete\/} $SU_5$ 
multiplets, see Ref.~\ref\munoz{C.~Munoz, hep-ph/0211066}.)

At three loops each soft $\beta$-function has many terms and 
some of the coefficients are quite large; given this 
it is worthwhile checking  whether perturbation
theory remains good in the MSSM. It is generally believed that the 
perturbation series for QFT $\beta$-functions are asymptotic in nature. 
The exact $\beta_g$ in the NSVZ scheme
\ref\jnsvz{D.R.T.~Jones, \plb123 (1983)  45 \semi 
V.~Novikov et al, \npb229 (1983) 381  \semi
V.~Novikov et al, \plb166 (1986) 329 \semi
M.~Shifman and A.~Vainstein, \npb277 (1986)  456}
for a pure (no matter) $N=1$ theory is clearly an exception, 
but in the presence of matter the 
perturbation series for $\gamma$ (and hence also the one for $\beta_g$) 
is probably asymptotic (for a discussion  
see Ref.~\ref\fjjb{P.M.~Ferreira, I.~Jack and D.R.T.~Jones,
\plb392 (1997) 376 
}). 
It is interesting that even for $n_5 = n_{10} = 0$ we find that, 
for the squark masses,   
three loop running corrections are typically  
larger than the two loop ones. We show explicitly how the 
three loop corrections affect the spectrum for some of the 
Snowmass benchmark points (SPS)
\ref\AllanachNJ{
B.C.~Allanach {\it et al.},
Eur.\ Phys.\ J.\ {\bf C} 25, 113 (2002)
}.

Another motive for the calculation and in particular for extending it to
$n_5, n_{10} \neq 0$ is that it enables us to explore the phenomenon of
``semi-perturbative unification'' as described by Kolda and
March-Russell (KMR)\ref\kmr{ C.F.~Kolda and J.~March-Russell, \prd 55 
(1997) 4252 }.  In this scenario, the gauge couplings increase at high
energies  but do not quite reach a Landau pole at gauge unification 
(contrasting with the non-perturbative unification of  Maiani et
al\ref\mpp{ L.~Maiani, G.~Parisi and R.~Petronzio,
\npb136 (1978) 115 (1978).
}, 
where the Landau pole occurs at $M_X$). It is  then
possible to argue  that there is a regime where 
perturbation theory remains reliable,  but  the
resulting physics differs markedly from that obtained  in the MSSM
case.\foot{For other related work on unification at strong coupling 
see 
Ref.~\ref\GhilenceaYR{
B.~Brahmachari, U.~Sarkar and K.~Sridhar,
Mod.\ Phys.\ Lett.\ A {\bf 8}  (1993) 3349\semi
R.~Hempfling,
\plb 351 (1995)  206\semi
K.S.~Babu and J.C.~Pati,
\plb 384 (1996)  140\semi
D.~Ghilencea, M.~Lanzagorta and G.G.~Ross,
\plb 415 (1997)  253 
\semi
G.~Amelino-Camelia, D.~Ghilencea and G.G.~Ross,
\npb 528 (1998)  35}}

Our calculations improve on those of KMR by including 
one loop threshold corrections and the complete 
three loop running corrections; we check that (for $n_5 = n_{10} = 0$) our 
results are consistent with those presented for the SPS points in 
Ref.~\AllanachJW\ and 
Ref.~\ref\GhodbaneKG{
N.~Ghodbane and H.U.~Martyn,
in {\it Proc. of the APS/DPF/DPB Summer Study 
on the Future of Particle Physics (Snowmass 2001) } ed. N.~Graf, hep-ph/0201233.
}.
While we provide more precise results, 
we support the conclusions of  KMR by demonstrating
that $n_5, n_{10} \neq 0$ can lead to
changes which are in some cases non-negligible, but are  
consistent with perturbation theory (modulo issues associated with 
the squark masses which we will discuss later), and can be readily 
distinguished from the MSSM.

\newsec{The Soft Beta functions}

The procedure for calculating the soft $\beta$ functions from 
$\beta_{g_i}$ and $\gamma$ is described in Ref.~\jjpa. The only 
subtlety relates to the $X$-function which arises in the soft scalar 
mass $\beta$-function; expressions for the leading and sub-leading 
contributions to this appear in Ref.~\jjpb.
Armed with these results it is straightforward 
to calculate the three-loop MSSM soft $\beta$-functions 
from the three-loop expressions for the $\beta_{g_i}$, $\gamma$ given 
in Ref.~\fjja. We have generalised this whole calculation to 
$n_5, n_{10}\neq 0$. 
The resulting expressions are very unwieldy; as an example 
we give the one, two and three loop results for $\beta_{m^2_{Q_t}}$, 
in the approximation that we retain only $g_3$ and the 
top quark Yukawa coupling $\lambda_t$ (in what follows 
we denote the third generation squarks as $Q_t, t^c, b^c$, 
the first or second generation squarks as 
$Q_u, u^c, d^c$, and we suppress $\lf$ loop factors. ):

\eqna\bqthree$$\eqalignno{
\beta_{m^2_{Q_t}}^{(1)} &= 
2 \lambda_t^2  (\Sigma_t +A_t^2)
-8(\frak{1}{60}g_1^2 M_1^2+\frak{3}{4} g_2^2  M_2^2+\frak{4}{3} g_3^2 M_3^2) 
& \bqthree a\cr
\beta_{m^2_{Q_t}}^{(2)} &= -20\lambda_t^4(\Sigma_t+2A_t^2)
+16g_3^4M_3^2(n_5+3n_{10}
-\frak{8}{3})\cr&
+\frak{16}{3}g_3^4 (2m^2_{Q_t}+m^2_{t^c}+m^2_{b^c}
+ (n_{10}+2)(m^2_{u^c}+2m^2_{Q_u})
+(n_5+2)m^2_{d^c}) & \bqthree b\cr
\beta_{m^2_{Q_t}}^{(3)} &=
[(1280k+\frak{20512}{9}+16n_5^2 +(\frak{6224}{9}+
\frak{320}{3}k)(n_5+3n_{10})\cr& + 96n_{10}n_5 +144n_{10}^2)M_3^2
+(\frak{320}{9}-\frak{16}{3}(n_5+3n_{10}))
(m^2_{t^c}+m^2_{b^c}+2m^2_{Q_t})\cr& +(2m^2_{Q_u}+m^2_{u^c})(\frak{640}{9}
-\frak{32}{3}n_5 + \frak{32}{9}n_{10}
     -\frak{16}{3}n_5n_{10}-16n_{10}^2)\cr&
+m^2_{d^c}(\frak{640}{9} +\frak{224}{9}n_5 -32n_{10} 
     -16n_5n_{10}-\frak{16}{3}n_{5}^2)] g_3^6\cr& 
-[(288+\frak{544}{3}k + 48(n_5+3n_{10}))M_3^2
-(192+\frak{1088}{9}k +32(n_5+3n_{10}))A_tM_3\cr&
+(\frak{272}{9}k+\frak{176}{3}+8(n_5+3n_{10}))(\Sigma_t+A_t^2)]\lambda_t^2g_3^4\cr&
+(\frak{160}{3}+32k)\left[M_3^2-2A_tM_3
+\Sigma_t
+2A_t^2)\right]\lambda_t^4g_3^2
+(6k+90)(\Sigma_t+3A_t^2)\lambda_t^6, & \bqthree c}$$
where $ {k = 6\zeta(3)}$, and $\Sigma_t = m^2_{Q_t}+m^2_{2}+m^2_{t^c}$.
For this special case, and also with $n_5 = n_{10} = 0$,  the three loop
result, Eq.~\bqthree{c}, was  given in  
Ref.~\ref\KazakovBT{D.I.~Kazakov, hep-ph/0208200}, 
except that in the corresponding expressions in this 
reference the squark  masses of different generations are not
clearly distinguished (as they must  be since the third generation evolves
differently from the other two).
Complete results for the three loop $\beta$-functions including 
all three gauge couplings and  $n_g\times n_g$ Yukawa matrices
may be obtained by application to the authors. 

Note that in our analysis we do not include ``tadpole'' contributions,
corresponding to renormalisation of the Fayet-Iliopoulos (FI) $D$-term. 
These contributions are not expressible exactly in terms of 
$\beta_{g_i}, \gamma$; for a discussion, and three loop results for the 
MSSM,  see 
Ref.~\ref\jjp{I.~Jack and D.R.T.~Jones, \plb473 (2000)\semi
I.~Jack, D.R.T.~Jones and S.~Parsons, \prd62 (2000) 125022\semi
I.~Jack and D.R.T.~Jones, \prd63 (2001) 075010}. For 
universal boundary conditions, the FI term is very small at 
low energies if it is zero at gauge unification; in this paper 
we restrict ourselves to universal boundary conditions
and ignore these contributions. 

\newsec{The Running Analysis}

In this section we examine the effect of the three loop corrections 
on the standard running analysis. We will focus on the standard treatment 
with universal boundary conditions at gauge unification, often 
termed CMSSM or MSUGRA. Thus we assume that at $M_X$ we 
have universal soft scalar masses ($m_0$), gaugino masses ($m_{\frak{1}{2}}$)
and $A$-parameters ($A$), and work in the third-generation-only 
Yukawa coupling approximation.  This is for ease of comparison with 
existing results rather than because we find the scenario particularly 
compelling. In the MSSM the corrections to the
dimensionless coupling  running analysis due to two  and three loop
corrections are comparatively small\fjja. As emphasised  by KMR, however,
this becomes less true for  $n_5, n_{10} \neq 0$. In particular, for
$n_5 + 3n_{10} = 6$, $\beta_{g_3} = 0$  at one loop, so that we need to 
consider at  least two loop corrections, and also 
three loops to verify that we remain within the perturbative domain.

We will present results 
as a function of $n_{10}$ (for $n_5 =0$); 
the dependence of physical quantities 
on $n_5$ (for $n_{10} = 0$) is qualitatively similar though not 
identical. 
As remarked by KMR, 
the mass scale of these additional multiplets being unknown it makes
sense to  parametrise their effects by taking $n_5, n_{10}$ to be
continuous variables. 
\smallskip
\epsfysize= 4in
\centerline{\epsfbox{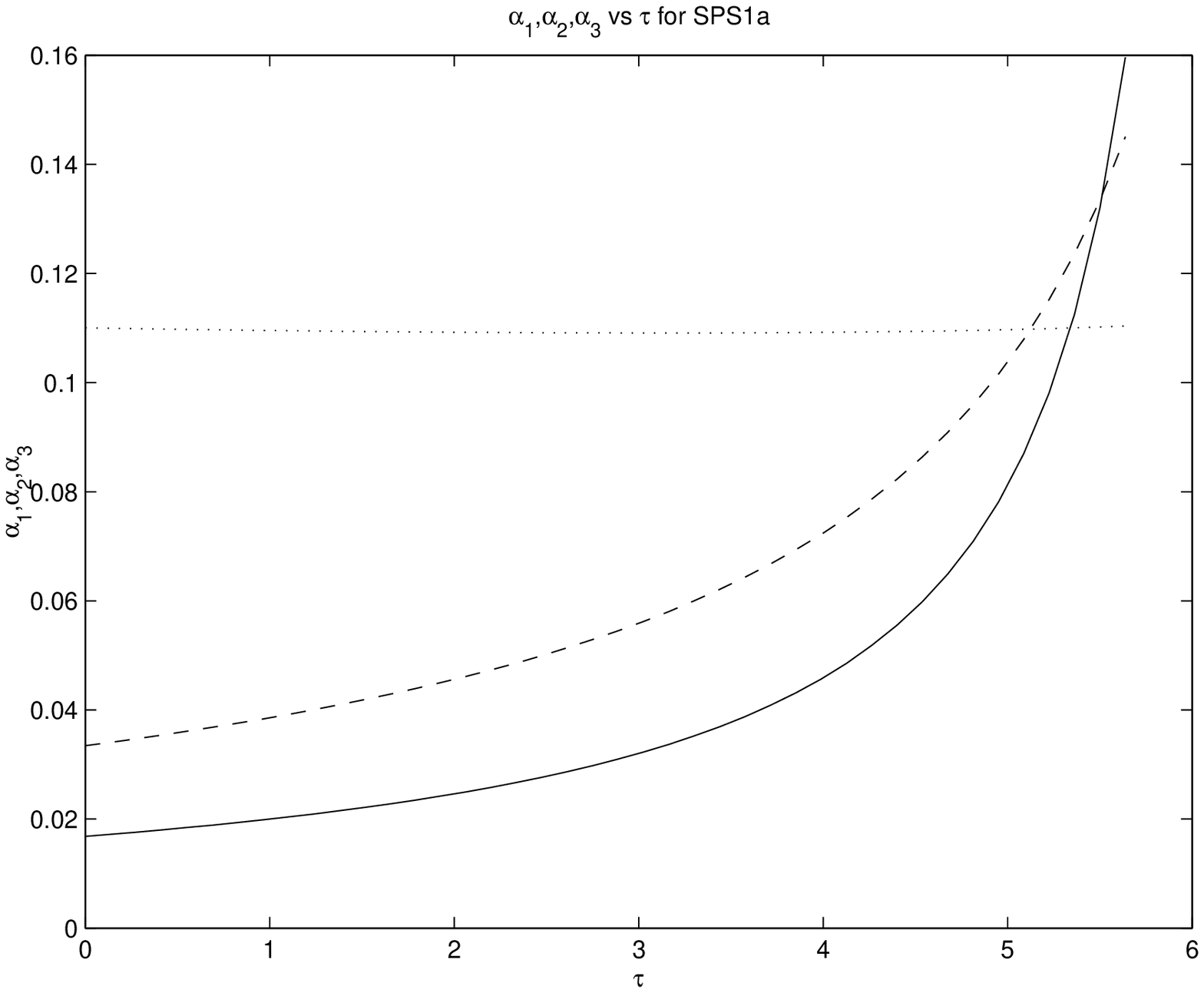}}
\inparg
{\it \noindent Fig.1:
Gauge coupling unification for $n_{10} = 1.7$.
Solid, dashed, and dotted lines correspond to 
$\alpha_1,\alpha_2,\alpha_3$ respectively.}
\medskip
\outparg

In Fig.~1 we show the evolution of the gauge couplings 
$\alpha_i = g_i^2/(4\pi)$ 
for $n_{10} = 1.7$, using three loop $\beta$-functions 
for all couplings. The couplings are plotted 
against $\tau = \frak{1}{2\pi}\ln(Q/M_Z)$; evidently we are still in the 
perturbative regime. The input parameters at $M_Z$ correspond 
to a typical supersymmetric mass spectrum;
specifically, the Benchmark point SPS1a, 
the details of which will be given later. 
We adjust input parameters according to 
the supersymmetric spectrum in order to account for 
threshold corrections in the manner of 
Ref.~\ref\pbmz{
D.M.~Pierce, J.A.~Bagger, K.T.~Matchev and R.J.~Zhang,
\npb491 (1997) 3 
}
\foot{In the first line of Eq.~37 of Ref.~\pbmz, 
the first term in the square bracket should read 
$-(m_{\ttilde_1}^2 
+ m_{\ttilde_2}^2)B_0 ( m_{\ttilde_2}, m_{\ttilde_1},0)$: 
i.e. it should have a minus sign. The corresponding exact result 
in Eq.~D49 is correct, however.}, and run up from $M_Z$ using the full \sic\ $\beta$-functions; 
thus the input 
values for the gauge couplings depend on the sparticle spectrum, and
are determined iteratively. We have set $n_{10} = 1.7$ because around 
this value  
the electroweak vacuum fails for the SPS1a input parameters; at this 
value it is interesting that (see Fig.~1)  
the (small) one-loop contribution to 
$\beta_{\alpha_3}$ is almost precisely cancelled by the 
two and three-loop ones. 
As already mentioned, if we have gauge coupling unification 
then this (and the scale at which it occurs) is
unaffected  by taking $n_5, n_{10}\neq 0$ in the one loop running
approximation.  This ceases to be true  for two and three loop running, 
and we must assume the existence of quite large GUT-scale threshold 
corrections to ensure unification. The 
unification scale corresponding to Fig.~1 (defined as where  $\alpha_1$
and $\alpha_2$ meet) is $M_U \approx 1\times 10^{17}\GeV$, 
significantly higher than in the MSSM. 

Turning to the soft parameters, note that  there are some large
coefficients in the expression for  $\beta_{m^2_{Q_t}}^{(3)}$ in
Eq.~\bqthree{c};  the coefficient of the $M_3^2g_3^6$  term, for
example is $O(10^4)$, even in the MSSM when $n_5 = n_{10} = 0$. For 
weakly-interacting particles and the gluinos, the three loop effects 
are quite small for zero or small $n_5, n_{10}$; but 
for the squark masses the three loop $\beta$-function coefficients 
are (at $M_Z$) typically (while smaller than the 
corresponding  1-loop coefficients) larger than the 
corresponding  two loop coefficients (even at $n_{5,10} = 0$) if 
$m_{\frak{1}{2}} > m_0$, as will be so, in fact, 
in the cases we shall present. 
One might well be tempted to interpret this as evidence for the 
asymptotic nature of the $\beta$-function series, as we 
mentioned in the Introduction.  
We will see the effects of this in the next section. 

As mentioned above, we adjust the input dimensionless 
parameters to accommodate threshold corrections\pbmz.  
We also incorporate the one-loop radiative corrections 
from this reference. For the input top mass at the weak scale 
we use\ref\bedn{A.~Bednyakov et al, 
Eur.\ Phys.\ J.\ {\bf C} 29 (2003) 87 }:
\eqn\topdef{\eqalign{
m_t(Q) & =  m_{t_{\rm pole}} \Bigl[ 1 - \frakk{\alpha_3}{3\pi}(5 - 3L)
- \alpha_3^2 \Bigl( 0.538-\frakk{43L}{24\pi^2}+\frakk{3L^2}{8\pi^2}\Bigr)\cr
&+O(\alpha_3^3) + \hbox {electroweak, sparticle contributions}\Bigr]}}
where $L = \ln [ {m_t(Q)^2}/Q^2]$.
This formula is identical to one from the up-to-date version of 
Allanach et al\AllanachJW.
 
For definiteness we analyse 
in detail particular  SPS benchmark points\AllanachNJ, 
and verify that our results 
(for $n_5 = n_{10} = 0$) are in accord with those obtained 
with existing computational tools\AllanachJW, \GhodbaneKG. 

\subsec {Benchmark point SPS 1a}

This point is a ``typical'' point in MSUGRA parameter space, 
with  $m_0 = 100\GeV$, $m_{\frak{1}{2}} = 250\GeV$, $A_0 = -100\GeV$,
$\tan\beta = 10$ and $\mu>0$. 
In Table~1 we  compare  our results for a selection of sparticle 
masses (at $n_5 = n_{10} = 0$) with the spread of results quoted in AKP
(note our convention that the predominantly
R-handed top squark is $\ttilde_2$).
\vskip3em
\vbox{
\begintable
 mass | 1loop  |2loops| 3loops | AKP \cr
 {\tilde g} | 630 | 615 |612 | 594-626 \cr
 \ttilde_2 | 404| 403| 395| 379-410\cr
 \utilde_L  |571 | 563 |555 | 536-570 \cr
\utilde_R | 551| 547| 538| 520-569 \cr
LSP | 105 | 97| 97|96.4-97.6
 \endtable}

\centerline{{\it Table~1:\/} Sparticle masses    
(in $\GeV$) for the SPS1a point}        
\medskip

We would expect our two loop results to correspond most
closely to AKP and we see that they are indeed consistent.
The effect of inclusion of three loop running is never greater than 
$2\%$; note, however,  that the shift 
caused by three loop running effects is comparable for $\utilde_{L}$ 
and larger for $\ttilde_2, \utilde_{R}$ than
that produced  by two loop running effects. 
In Fig.~2 we plot the ratio
of the light  stop mass to the gluino mass as a function of $n_{10}$
(for $n_5 = 0$),  and we see that both two and three loop effects
increase dramatically  as $n_{10}$ increases. (NB the increase 
in the squark/gluino mass ratio with $n_{5,10}$ 
observed by KMR applies to the squarks of the first two generations). 

In Fig.~3 we plot the ratio of the LSP mass to the gluino mass 
as a function of $n_{10}$ (for $n_5 = 0$). Here we see that the impact of 
the three loop running corrections is less marked but still appreciable at 
large $n_{10}$. 

\epsfysize= 3in
\centerline{\epsfbox{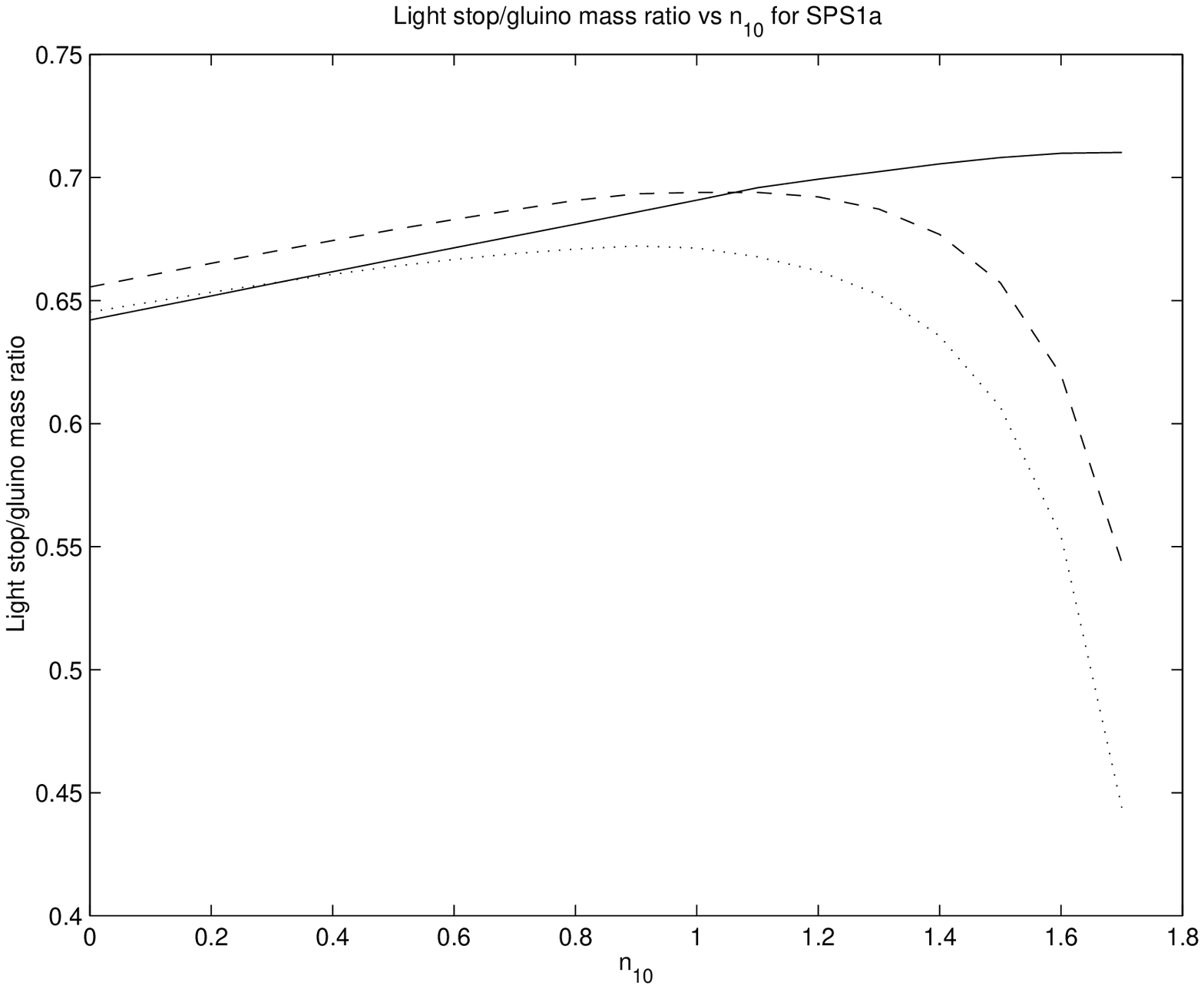}}

\inparg
{\it \noindent Fig.2:
Plot of the light stop/gluino mass ratio against $n_{10}$
for SPS1a.
Solid, dashed and dotted lines correspond to one, two and 
three loop running respectively. 
}
\medskip
\outparg

\smallskip
\epsfysize= 3.0in
\centerline{\epsfbox{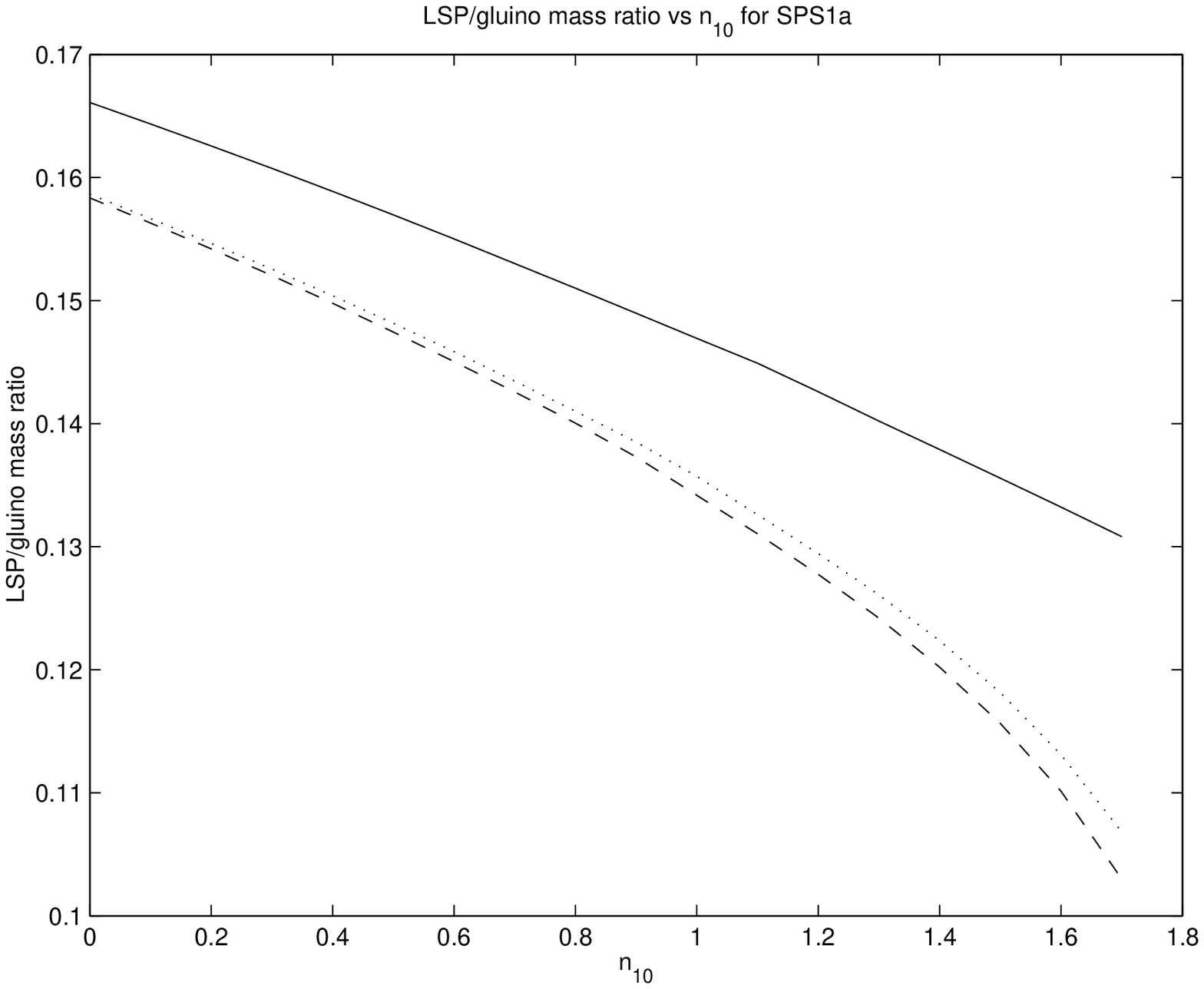}}
\inparg
{\it \noindent Fig.3:
Plot of the LSP/gluino mass ratio against $n_{10}$ for SPS1a.
Solid, dashed and dotted lines correspond to one, two and
three loop running respectively.
}
\smallskip
\outparg

%
%
%

\subsec {Benchmark point SPS 5}

This point differs from the previous one in having a large 
value of the $A$-parameter: 
with  $m_0 = 150\GeV$, $m_{\frak{1}{2}} = 300\GeV$, $A_0 = -1\TeV$,  
$\tan\beta = 5$ and $\mu>0$.
The  contributions of $\mu, A_0$   
to the off-diagonal term in the stop mass matrix have the same sign,
and the magnitude of $A_0$ is large, resulting in a light stop.
For this point we obtain in the MSSM  
(with $n_5 = n_{10} = 0$) the results shown in Table~2.

\vskip3em
\vbox{
\begintable
 mass | 1loop  |2loops| 3loops | AKP \cr
 {\tilde g} | 745 | 731 |729 | 705-730 \cr
 \ttilde_2 | 236 | 250| 231| 232-248 \cr
 \utilde_L  |682| 674 |666| 642-681 \cr
\utilde_R | 657| 655| 645| 622-681 \cr
LSP | 128| 120| 120 |118.7-121.1
 \endtable}

\inparg 
\centerline{{\it Table~2:\/} Sparticle masses    
(in $\GeV$) for the SPS5 point}        
\medskip
\outparg  
Once again the inclusion of three loop effects causes mass shifts 
of less than $2\%$, 
except for the light stop where 
there is an effect of about $8\%$. 
The light stop mass comes from a $2\times 2$ matrix with 
large off-diagonal entries; 
the large three loop shift is caused essentially by the changes 
in all the  entries due  to 
the comparatively large contributions to the three loop 
soft $\beta$-functions, as we described earlier.
The mass of the light stop is also very sensitive indeed to the input 
$m_t(M_Z)$, which in turn depends on the sparticle spectrum and the 
input top pole mass (which we have taken to be $174.3\GeV$). 
However (for a given $m_{t_{\rm pole}}$), 
$m_t(M_Z)$ changes very little when we include the 3 loop corrections.
Note that our two loop result is slightly above 
the range obtained by AKP.
It is also  very sensitive to $n_5, n_{10}$;
in Figure 4 we again plot the  light stop/gluino mass ratio against $n_{10}$:
this time the electroweak vacuum fails for $n_{10}\approx 0.3$. 
Remarkably, the three loop and 
two loop corrections cancel almost exactly near $n_{10}=0$, while at 
$n_{10}\sim 0.3$ the two loop corrections are rather small. 

\vskip 1cm
\epsfysize= 3in
\centerline{\epsfbox{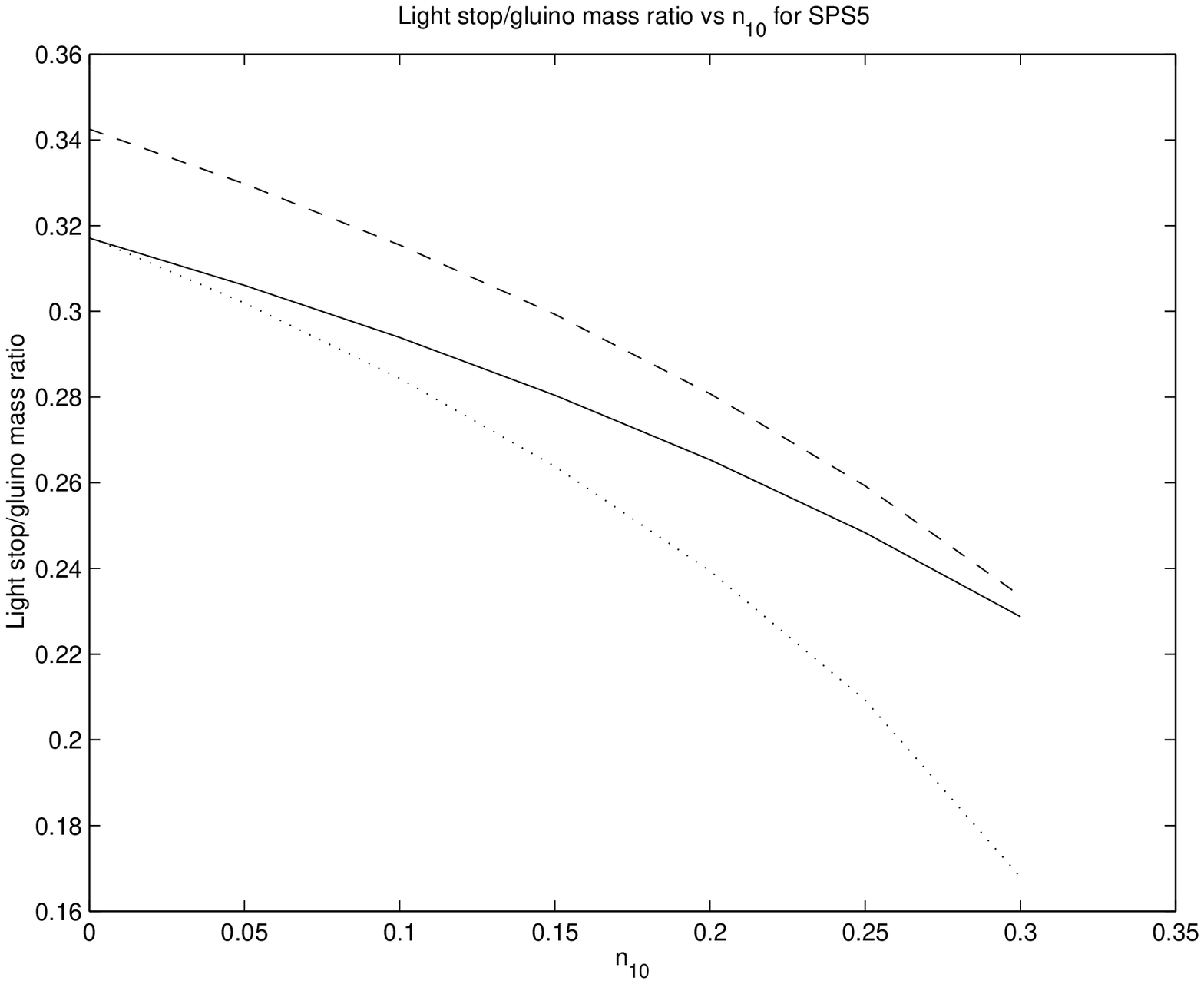}}

\inparg
{\it \noindent Fig.4:
Plot of the light stop/gluino mass ratio against $n_{10}$
for SPS5. Solid, dashed and dotted lines correspond to one, two and
three loop running respectively.}
\medskip
\outparg

\smallskip
\epsfysize= 3.0in
\centerline{\epsfbox{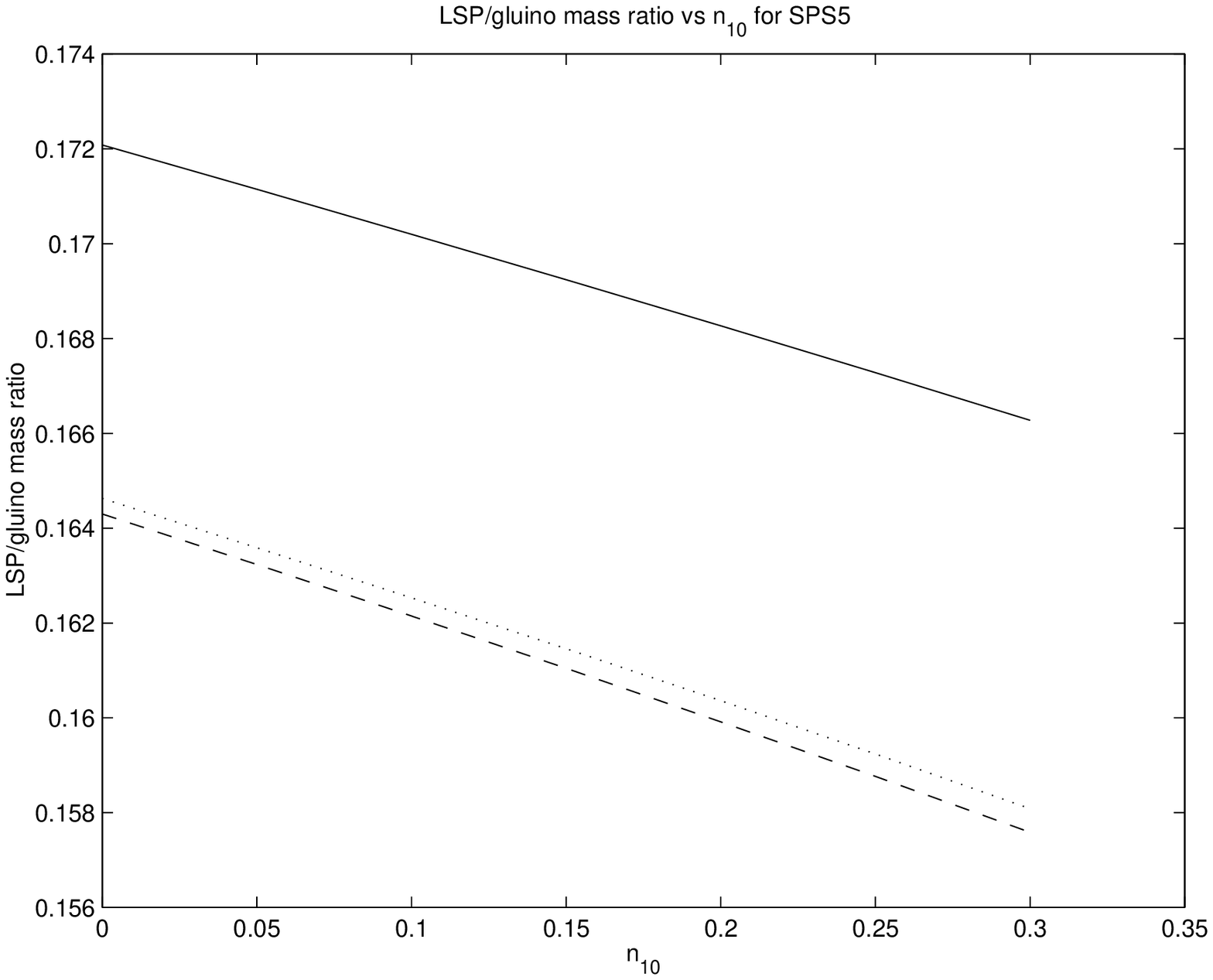}}

\inparg
{\it \noindent Fig.5:
Plot of the LSP/gluino mass ratio against $n_{10}$ for SPS5.
Solid, dashed and dotted lines correspond to one, two and
three loop running respectively.}
\smallskip
\outparg

In Fig.~5 we plot the ratio of the LSP mass to the gluino mass for SPS5
as a function of $n_{10}$ (for $n_5 = 0$). Here we see that the impact of       
the three loop running corrections is less marked but still appreciable 
as $n_{10}$ increases.

\newsec{Conclusions}

We have extended typical detailed running coupling analyses for  a
selection of  MSSM SPS benchmark points to incorporate three loop
$\beta$-function corrections for the running masses and couplings.
Generally speaking the effect of the three loop running corrections is
at most $2\%$ and of the same size or 
smaller than that of the two loop corrections, except for 
squark masses where it can be larger; simply because the 
three-loop $\beta$-function  coefficients are larger than the two-loop ones. 
For the light stop mass for the SPS5 point, 
we see an $8\%$ effect; this happens because this 
mass results from the diagonalisation of a matrix 
with large off diagonal entries which all change as described above.
We have  also performed the same analysis for
the MSSM extended to incorporate  additional matter in the form 
of $SU_5$ $5$ and $10$ representations.  As the amount of such matter 
is increased the effect of two and three loop corrections becomes 
more dramatic, as the one-loop $\beta$-function  for $\alpha_3$ decreases.

Given some detail of the sparticle spectrum, it will be comparatively
easy  to distinguish, for example, the CMSSM and AMSB scenarios; however
in the context of the former, disentangling the possible impact of 
additional matter and the effect of radiative corrections will be more
difficult. 

One may expect in general that 
two loop threshold/pole mass corrections will be competitive with 
the three loop running corrections that we have described, and 
so for accurate predictions one should include both. At this level 
one is also  sensitive to the experimental uncertainty 
in $m_t$ (for the light stop sometimes very sensitive, as we have 
described) and the strong coupling $\alpha_3 (M_Z)$. 
It is feasible that by the time sparticles are discovered 
complete two loop threshold corrections will be available, and
that both these uncertainties will also be reduced, so that 
significantly more accurate sparticle spectrum predictions 
will be possible. It appears, however,  
from the apparently asymptotic nature of the 
squark mass $\beta$-functions that squark mass predictions 
with an accuracy greater than around $2\%$ will not be possible
using perturbation theory.

\medskip

\bigskip\centerline{{\bf Acknowledgements}}\nobreak

DRTJ was  supported by a PPARC Senior Fellowship,  and was visiting the
Aspen Center for Physics while part of this work was done. AK was
supported  by an Iranian Government Studentship.  We thank Ben Allanach, 
Jon Bagger, Ruth Browne and John Gracey for conversations.   
 

\listrefs
\bye